\documentclass{iaus}
\usepackage{graphicx}

\title[Evidence for Shock-Shock Interaction in the Jet of CTA\,102] %% give here short title %%
{Evidence for Shock-Shock Interaction in the Jet of CTA\,102}

\author[C. M. Fromm et al.]
{C. M. Fromm$^1$, M. Perucho$^2$, T. Savolainen$^1$, E. Ros$^{2,1}$, A.~P.~Lobanov$^1$  J. A. Zensus$^1$ \and A. L\"ahteenm\"aki$^3$} 

\affiliation{$^1$Max-Planck-Institut f\"ur Radioastronomie, \\ Auf dem H\"ugel 69,
D-53121 Bonn, Germany\\ email: {\tt cfromm@mpifr.de} \\[\affilskip]
$^2$Departament d'Astronomia i Astrof\'{\i}sica, Universitat de Val\`encia, \\  E-46100, Burjassot, Val\`encia, Spain  \\[\affilskip]
$^3$ Aalto University, Mets\"ahovi Radio Observatory, \\
FI-02540 Kylm\"al\"a, Finland}

\pubyear{2010}
\volume{275}  %% insert here IAU Symposium No.
\pagerange{119--120}
\jname{Jets at all Scales}
\editors{Gustavo E. Romero, Rashid A. Sunyaev \and Tomaso Belloni}
\begin{document}

\maketitle

\begin{abstract}
We have found evidence for interaction between a standing and a traveling shock in the jet of the blazar CTA~102.  Our result is based in the study of the spectral evolution of the turnover frequency-turnover flux density ($\nu_m ,S_m$) plane.  The radio/mm light curves were taken during a major radio outburst in April 2006.
\keywords{galaxies: active, -- galaxies: jets, -- radio continuum: galaxies, -- radiation mechanisms: non-thermal, -- galaxies: quasars: individual: CTA\,102}
\end{abstract}

\firstsection % if your document starts with a section,
\section{Introduction}
The blazar CTA\,102 (z=1.037) shows a curved jet which exhibits apparent velocities up to 15.4\,c (\cite[{Lister} et al. 2009]{Lis09}). The analysis of single-dish light curves and 43\,$\mathrm{GHz}$ VLBI observations of \cite[{Hovatta} et al. (2009)]{Hov09} and \cite[{Jorstad} et al. (2005)]{Jor05} yields bulk Lorentz factors, $\Gamma$ between 15 and 17 and Doppler factors, $\delta$ between 15 and 22. These results lead to the picture that CTA\,102 harbors a highly relativistic jet. The source underwent a major radio flare from millimetre to centimetre wavelength during spring 2006. We have analyzed the single-dish data available for this source and here we present results of our modeling of the spectral evolution during the radio flare.

\section{Single-Dish Light Curves}
For our analysis of the 2006 radio flare in CTA\,102 we used radio/mm light curves spanning from 4.8\,$\mathrm{GHz}$ to 340\,$\mathrm{GHz}$. In order to obtain simultaneous spectra, we interpolated the observed light curves and performed a spectral analysis. Within this analysis we subtracted a quiescent spectrum from the obtained data and fitted a self-absorbed spectrum, defined by $S(\nu)=C\left(\nu/\nu_1\right)^{\alpha_t}\left\{1-\exp\left[-\left(\nu/\nu_1\right)^{\alpha_0-\alpha_t}\right]\right\}$, where $S(\nu)$ is the flux density, $\nu_1$ is the frequency at which the opacity $\tau_s=1$, and $\alpha_t$ and $\alpha_0$ are the spectral indices for the optically thick and optically thin parts of the spectrum, respectively. The turnover frequency, $\nu_m$, and the turnover flux density, $S_m$, can be calculated from the first and the second derivative of the synchrotron spectrum and they can be regarded as the characteristics of the spectrum. The uncertainties on the derived spectral values were obtained by performing Monte Carlo simulations. The derived spectral evolution showed a discrepancies from the standard evolution with in the shock-in-jet model (\cite[{Marscher} \& {Gear} 1985]{Mar85}), visible as a second hump in the $\left(\nu_m-S_m\right)$ plane (\cite[{Fromm} 2009]{Fro09}).

\section{Modeling}
The shock-in-jet model assumes a power law relation between the turnover frequency, $\nu_m$, and the turnover flux density, $S_m$, where the exponent depends on the dominant energy loss mechanism (Compton, Synchrotron  and Adiabatic) and the evolution of the physical properties in the jet. The model assumes a power law distribution for the relativistic electrons, $N(E)\propto KE^{-s}$ and the evolution of the constant $K$ follows the relation $K\propto R^{-k}$, where $R$ is the radius of the jet. Furthermore the  the evolution of the magnetic field can be expressed as $B\propto R^{-b}$, the evolution of the Doppler factor is $\delta\propto R^{-d}$ and of the jet radius, $R\propto L^{r}$, where $L$ corresponds to the distance along the jet. Applying this model to the derived spectral behaviour leads to the following results for the evolution of the physical quantities. 

\begin{table}[h!]
\centering
\small{
\begin{tabular}{|c| c| c| c| c| c| c|}
\hline
time& stage &b & s & k & d & r \\
\hline
 2005.6 - 2005.9&Comp/Adi &  1.98 & 2.77 & 1.89 & 0.11 & 0.70\\
2005.9 - 2006.3 &Comp&$ -$3.58 & 1.20 & 2.13 & 1.58 & $-$0.19 \\
2006.3 - 2006.5 & Adi&1.89 & 2.64 & 2.33 & $-$0.22 &1.00 \\
2006.5 - 2006.8 & Adi&1.89 & 2.64 & 3.05 & $-$0.11  &1.00 \\
\hline
\end{tabular}}
\end{table}

\section{Discussion}
The evolution of the parameter $r$ (jet opening) could be an indication for a non-pressure matched jet and the behaviour of the power law index $s$ could be interpreted as a re-acceleration of relativistic particles. Together with the variation in the magnetic field $(b)$ and the Doppler factor $(d)$ our result could be associated with a traveling-standing shock interaction. Standing shocks are common features in non-pressure matched jets. At the position of the standing shock, the traveling shock will encounter a region of  increased particle density and magnetic field strength and will generate, due to shock acceleration at the shock front, a higher emissivity (\cite[{G\'omez} et al. 1997]{Gom97}). This increase in the emissivity could lead to the observed spectral behaviour. A more detailed discussion will be presented elsewhere. Future work should include the analysis of available multi-frequency VLBI observations in combination with relativistic magneto-hyrdodynamic simulations which could confirm our scenario of a shock-shock interaction in CTA\,102.

\end{document}